\documentclass[twocolumn,english,aps,prl,superscriptaddress,notitlepage,longbibliography]{revtex4-2}
\usepackage[latin9]{inputenc}
\usepackage{xcolor}
\definecolor{document_fontcolor}{rgb}{0, 0, 0}
\color{document_fontcolor}
\usepackage{babel}
\usepackage{mathtools}
\usepackage{amsmath}
\usepackage{amssymb}
\usepackage{graphicx}
\usepackage{seqsplit}
\usepackage[pdfusetitle,bookmarks=true,bookmarksnumbered=false,bookmarksopen=false,breaklinks=false,pdfborder={0 0 1},backref=false,colorlinks=true]{hyperref}
\hypersetup{urlcolor=blue,citecolor=blue,hyperfootnotes=blue,linkcolor=blue}

\begin{document}

\title{Langevin Theory of Non-Markovian Quantum Dynamics:\\ Application to Delayed Coherent Feedback and the Laser Linewidth}

\author{Marc Cuenca-Lar\`as}
\affiliation{Departament d'\`Optica i Optometria i Ci\`encies de la Visi\'o, Universitat de Val\`encia, Av. Vicent Andr\'es Estell\'es 19, 46100 Burjassot, Spain}

\author{Ming Li}
\affiliation{Shenzhen Institute for Quantum Science and Engineering, Southern University of Science and Technology, Shenzhen 518055, China}

\author{Carlos Navarrete-Benlloch}
\affiliation{Departament d'\`Optica i Optometria i Ci\`encies de la Visi\'o, Universitat de Val\`encia, Av. Vicent Andr\'es Estell\'es 19, 46100 Burjassot, Spain}

\author{Germ\'an J. de Valc\'arcel}
\affiliation{Departament d'\`Optica i Optometria i Ci\`encies de la Visi\'o, Universitat de Val\`encia, Av. Vicent Andr\'es Estell\'es 19, 46100 Burjassot, Spain}

\begin{abstract}
Phase-space methods are powerful tools for the treatment of Markovian open quantum systems: they map the reduced dynamics of a system S, in interaction with an environment E, exactly onto Langevin equations for c-number stochastic variables, as opposed to Heisenberg--Langevin equations for operators. Langevin equations provide analytical insight in key regimes and excel at handling strong nonlinearities and couplings, where other methods often falter. Extending phase-space methods to non-Markovian dynamics, however, has remained a long-standing challenge. Here we address this gap by applying phase-space representations to the full S+E system; integrating out the environmental degrees of freedom then yields a general Langevin framework for S that incorporates both deterministic and stochastic contributions from E. Normally ordered representations---such as the Glauber--Sudarshan $P$ representation and its positive variant due to Drummond and Gardiner---lead to Langevin equations in which (i) non-Markovian effects emerge exclusively in the deterministic terms, via a memory kernel, and (ii) noise contributions vanish when E is initially in the vacuum state. As a demonstration of the power of this framework, we address the paradigmatic problem of delayed coherent feedback, in which the system is driven by its own state a time $T$ in the past, and study its impact on the laser linewidth: we recover the narrowing observed well above threshold and predict an enhanced narrowing just above it. Crucially, the number of stochastic variables scales linearly with the system size, making the framework suitable for problems ranging from a few degrees of freedom to genuinely many-body systems. This opens the way to the systematic study of non-Markovian driven-dissipative quantum systems using the same analytical and numerical tools that have long made phase-space methods so successful in the Markovian regime.

\end{abstract}

\maketitle

\emph{Introduction.---}The theory of open quantum systems (OQS) \cite{BreuerPetruccioneBook,GardinerZollerBook} has been a cornerstone of quantum technological advances in recent decades \cite{QTbook1,QTbook2,DowlingBook21,DowlingBook13}. In its memoryless, or Markovian, formulation \cite{BreuerPetruccioneBook,GardinerZollerBook,CarmichaelBook,Manzano20,CNB-QOnotes},
a system S interacts with a large environment E of negligibly short memory, as E relaxes on timescales much shorter than any of the system's. Consequently, the \emph{change} in the system's state at any instant depends only on its present state, not on its history. This assumption, however, is overly restrictive and breaks down in many relevant problems, from quantum Brownian motion \cite{Ferialdi17} to systems interacting with structured environments \cite{deVega08,CNB11,Krinner18,Liu11,deVega17,Li19b} or processes such as coherent feedback \cite{Lloyd00}, which have gained experimental significance in recent years. A variety of approaches have emerged to address these non-Markovian scenarios \cite{BreuerRev16,Li18,Li19a,Li19b,Rivas14,Shrikant23}, including generalized \cite{Vacchini16}, stochastic \cite{Stockburger02}, and nonlinear \cite{Degenfeld15} master equations, tensor-network methods \cite{Finsterholz20,Strathearn18}, and collisional models \cite{Campbell18,Ciccarello18,Ciccarello21}. Special attention has been devoted to systems with delayed coherent feedback, for which several methods have been developed, including those based on tensor networks \cite{Pichler16,Pichler17,Guimond17,Vodenkova24}, generalized Heisenberg--Langevin equations \cite{Grimsmo15,Kraft16,Nemet16}, Schr\"{o}dinger-picture maps \cite{Grimsmo15,Whalen17}, and quantum trajectories \cite{Crowder20,Finsterholz20}.

In quantum optics and related fields, the efficient simulation of OQS dynamics often rests on phase-space techniques (PST), such as the Wigner and Glauber--Sudarshan representations \cite{CarmichaelBook,CarmichaelBook2}. These methods become indispensable when nonlinearities play a significant role or when the number of excitations is large \cite{Gilchrist97,CarmichaelBook,CarmichaelBook2}, since in such cases the Hilbert space can become numerically intractable for brute-force methods. The dynamics of Markovian OQS is typically described by a
Lindblad master equation for the reduced state (density operator), which phase-space techniques can map onto a set of c-number stochastic (Langevin) equations \cite{CarmichaelBook,CarmichaelBook2}. This approach is highly advantageous because Langevin equations \cite{CarmichaelBook,CarmichaelBook2,GardinerZollerBook} (i) resemble classical dynamical equations augmented with noise, aiding physical intuition; (ii) provide an exact model of observable dynamics, which can be efficiently simulated numerically; (iii) offer a tractable
method for studying phase dynamics \cite{CNB08,CNB10,CNB17combs,CNB17}; and (iv) scale linearly with the number of modes or system size, enabling simulations of complex many-body systems \cite{DrummondDeuar07,Drummond16rev,Kiesewetter22,Kiesewetter22b,Deuar21,Deuar21b,Dellios23,Orso25}. In contrast, the dynamics of non-Markovian OQS is described, at best, by generalized master equations \cite{Vacchini16}, whose nonlocality in time complicates their direct mapping onto Langevin equations.

In this work, we extend PST to non-Markovian quantum systems, obtaining c-number Langevin equations for the system S alone after formally integrating out the environment E. These stochastic equations enable both analytical treatment and efficient numerical simulation with standard methods. Our approach applies PST directly to the full S+E system, using normally ordered representations such as the Glauber--Sudarshan $P$ representation. We show that the resulting Langevin equations for S encode the environment's influence
through (i) a tunable memory kernel that controls the non-Markovian dynamics, and (ii) an additive noise that depends on the initial state of the environment but that, remarkably, vanishes for a vacuum state.

As a first application of the framework, we consider the problem of delayed coherent feedback. We show that our theory yields the simplest conceivable model, in which feedback enters exactly as it would classically, with no extra quantum noise---at variance with previous studies \cite{Kraft16,Nemet16}. Moreover, while these prior treatments required a linearized description of quantum fluctuations, our method allows for the numerical simulation of the full quantum dynamics, exact in principle. We then study the paradigmatic and technologically relevant case of a laser under delayed coherent feedback. Specifically, we analyze the laser linewidth, reproducing known results well above threshold and offering new predictions elsewhere in parameter space.

\emph{Stochastic equations for a generic environment.---}We illustrate our approach in the archetypal case \cite{GardinerZollerBook,CarmichaelBook,CNB-QOnotes,deVega17} of a single-mode bosonic quantum system S coupled to a multimode bosonic environment E via a particle-conserving bilinear Hamiltonian, which we treat without approximations. In addition, S can be coupled to other subsystems and baths that lead, e.g., to dissipation or pumping. We denote by $\hat{a}$ and $\hat{b}_j$ the annihilation operators of S and of the $j$-th mode of E, respectively, which obey canonical commutation relations $[\hat{a},\hat{a}^\dagger]=1$ and $[\hat{b}_j,\hat{b}_l^\dagger]=\delta_{jl}$, with any other commutator vanishing. We assume a dynamics governed by a Liouville equation for the joint density operator $\hat{\rho}$ of the S+E system,
\begin{equation}
\frac{d\hat{\rho}}{dt}=\bigg[ \frac{\hat{H}}{\mathrm{i}\hbar},\hat{\rho}\bigg]+\mathcal{L}_\text{S}[\hat{\rho}],\label{Liouville}
\end{equation}
where $\hat{H}=\hat{H}_\textrm{S}+\hat{H}_\text{E}+\hat{H}_\text{SE}$ contains the free evolution of S and E, $\hat{H}_\text{S}=\hbar\omega_\text{S}\hat{a}^\dagger\hat{a}$ and $\hat{H}_\text{E}=\sum_j\hbar\omega_j\hat{b}_j^\dagger\hat{b}_j$, and their mutual interaction, $\hat{H}_\text{SE}=\sum_j\hbar(g_j\hat{b}_j\hat{a}^\dagger+g_j^\ast\hat{b}_j^\dagger\hat{a})$, where $\{g_j\}_j$ are complex coupling constants that set the interaction strengths. The Liouville superoperator $\mathcal{L}_\text{S}$ accounts for any dynamics of S not generated by $\hat{H}$, arising from other Hamiltonian and irreversible processes such as nonlinearities, coupling to other subsystems, and exchange (driving, dissipation, pumping, etc.) with other baths. 

Our approach uses any suitable normally ordered phase-space representation for $\hat{\rho}$ \cite{GardinerZollerBook,CarmichaelBook,CarmichaelBook2,WallsMilburnBook}, turning the Liouville equation (\ref{Liouville}) into a Fokker--Planck equation, which is in turn equivalent to a set of Langevin equations for c-number stochastic variables via It\^o's lemma \cite{GardinerBook}. We then integrate out the environmental degrees of freedom, obtaining stochastic equations for S only. Note that this is not the usual approach: for Markovian systems, one first finds the quantum master equation for the reduced density operator of S alone, i.e., after the environmental degrees of freedom have been traced out from the von Neumann equation in second-order perturbation theory \cite{GardinerZollerBook,CarmichaelBook,BreuerPetruccioneBook,WallsMilburnBook,Manzano20,CNB-QOnotes}, and then applies any suitable phase-space representation \cite{Drummond80,GardinerZollerBook,CarmichaelBook,CarmichaelBook2,WallsMilburnBook}. In the following we use the simplest normally ordered representation, the so-called Glauber--Sudarshan $P$ representation \cite{Glauber,Sudarshan}, but our approach applies directly to more sophisticated representations, such as the positive $P$ representation by Drummond and Gardiner \cite{Drummond80}.

Upon applying the $P$ representation to Eq.~(\ref{Liouville}), the operators $\hat{a}$, $\hat{a}^\dagger$, $\hat{b}_j$, and $\hat{b}_j^\dagger$ become equivalent, in normal order, to the c-number variables $\alpha$, $\alpha^\ast$, $\beta_j$, and $\beta_j^\ast$, which obey the following Langevin equations \cite{SupMat}:
\begin{subequations}\label{StochasticSE} 
\begin{align}
\dot{\alpha} & =-\mathrm{i}\omega_\text{S}\alpha-\mathrm{i}\sum_j g_j\beta_j+\mathrm{L}_\text{S}(\alpha,\alpha^\ast),\label{dalpha0}
\\
\dot{\beta}_j & =-\mathrm{i}\omega_j\beta_j-\mathrm{i}g_j^\ast\alpha,\label{dbeta}
\end{align}
\end{subequations}
where the dot denotes $d/dt$. The equivalence, which is exact, means that $\langle\hat{a}^{\dagger m}\hat{a}^{n}\otimes_j\hat{b}_j^{\dagger m_j}\hat{b}_j^{n_j}\rangle=\overline{\alpha^{*m}\alpha^{n}\prod_j\beta_j^{*m_j}\beta_j^{n_j}}$, i.e., quantum expectation values (brackets) of \emph{normally ordered} operators equal the corresponding stochastic averages (overbars). The function $\mathrm{L}_\text{S}(\alpha,\alpha^\ast)$ in Eq.~(\ref{dalpha0}) gathers all terms coming exclusively from $\mathcal{L}_\text{S}[\hat{\rho}]$ in Eq.~(\ref{Liouville}). It typically contains damping, nonlinearities, and noise terms but, importantly, $\mathrm{L}_\text{S}$ does not depend on E, i.e., on the set $\{\beta_j,\beta_j^\ast\}_j$; an example is given below for the laser.

Since our goal is a description of the dynamics of S alone, we formally integrate Eq.~(\ref{dbeta}) as $\beta_{j}(t)=\beta_{j}(0)\mathrm{e}^{-\mathrm{i}\omega_{j}t}-\mathrm{i}g_{j}^{\ast}\int_{0}^{t}dt^{\prime}\mathrm{e}^{-\mathrm{i}\omega_{j}(t-t')}\alpha(t^{\prime})$, and insert the result into Eq.~(\ref{dalpha0}). Introducing the \textit{memory time} $\tau\coloneqq t-t'$ and the slowly varying interaction-picture amplitude $z(t)\coloneqq\alpha(t)\mathrm{e}^{\mathrm{i}\omega_\text{S}t}$, we are left with
\begin{subequations}\label{Lange0} 
\begin{align}
\dot{z}(t) & =\mathcal{E}[z;t]+L_\text{S}(z,z^\ast),\label{dzdt}
\\
\mathcal{E}[z;t] & \coloneqq-\int_0^t d\tau K(\tau)z(t-\tau)+\Gamma(t),\label{def-E}
\end{align}
\end{subequations}
where $L_\text{S}=\mathrm{e}^{\mathrm{i}\omega_\text{S}t}\mathrm{L}_\text{S}$
under the change of variables. The memory kernel, defined as
\begin{equation}
K(\tau)\coloneqq\sum_j\vert g_j\vert^2 \mathrm{e}^{\mathrm{i}(\omega_\text{S}-\omega_{j})\tau},\label{def-K}
\end{equation}
is a Hermitian function of $\tau$, i.e., $K(-\tau)=K^\ast(\tau)$, and dictates how the environment deterministically drives the system in response to their mutual interaction. Note that $K(0)\ge|K(\tau)|$ $\forall\tau$. The term $\Gamma(t)\coloneqq-\mathrm{i}\sum_j g_j\beta_j(0)\mathrm{e}^{\mathrm{i}(\omega_\text{S}-\omega_j)t}$ is a stochastic driving that depends on the initial environmental state through the set of initial values $\{\beta_{j}(0)\}_j$. Note that this is just an additive noise if one further assumes, as is commonplace in the theory of open systems \cite{GardinerZollerBook,CarmichaelBook,BreuerPetruccioneBook,Manzano20,CNB-QOnotes}, that at $t=0$ the system and the environment are uncorrelated.

A main result is already evident from Eqs. (\ref{Lange0}): the effects of E on S, encapsulated in $\mathcal{E}[z;t]$, simply add to the dynamics $L_\text{S}$ that S would have if decoupled from E. Regarding $\Gamma(t)$, although it is reminiscent of the input operators obtained in the Heisenberg picture when integrating out the environment \cite{GardinerZollerBook,CNB-QOnotes,WallsMilburnBook}, the situation is far simpler in our approach: for any normally ordered representation---such as the $P$ representation used here---$\Gamma(t)$ vanishes if the environment starts in the vacuum state, implying $\beta_{j}(0)=0$ for all $j$ \cite{SupMat}. Hence, under this common assumption, which we adopt in what follows, $\Gamma(t)=0$---there is no added noise due to E---and all the information about the environment is contained in the memory kernel $K$, irrespective of its Markovian or non-Markovian nature. In particular, under the Markov approximation, one assumes that $K(\tau)$ decays very quickly away from $\tau=0$ compared to the system's evolution rate, idealizing it as a delta function, i.e., $K(\tau)=2\gamma_{0}\delta(\tau)$, with $\gamma_{0}$ a real and positive constant: the environmental term (\ref{def-E}) then reduces to $\mathcal{E}[z;t]=-\gamma_{0}z(t)$, recovering the familiar irreversible decay into E.

\emph{The memory kernel: Application to delayed coherent feedback}.---We now apply the formalism to the generic problem of delayed coherent feedback. Rather than proposing a specific model for the microscopic parameters $\{\omega_{j},g_{j}\}_{j}$, our framework allows us to set the memory kernel on physical grounds. Denoting the delay by $T$, the kernel $K(\tau)$ must be strongly peaked around $\tau=T$, a peak we idealize as a Dirac delta, under the assumption that the system's evolution is negligible on the time scales of the actual peaks widths. Since $K$ must be Hermitian it must also be peaked around $\tau=-T$, i.e., in the future. Nevertheless this poses no causality problem, as the values of $K(\tau<0)$ do not contribute to the dynamics; see Eqs. (\ref{Lange0}). In addition, since the kernel must be maximum at $\tau=0$, it must have a peak at that value as well, whose weight relative to the $\tau=\pm T$ peaks is proven to be 2 on physical grounds \cite{SupMat}. We finally obtain
\begin{equation}
K(\tau)=f\left[2\delta(\tau)-{e^{\mathrm{i}\phi}}\delta(\tau-T)-{e^{-\mathrm{i}\phi}}\delta(\tau+T)\right],\label{K_choice}
\end{equation}
where $f\ge0$ is a rate characterizing the feedback strength and $\phi$ the phase accumulated by the feedback (equal to the oscillatory phase accumulated during a delay time, $\omega_{\text{S}}T$, plus any additional phase imprinted in the feedback loop). This coincides with the kernel found, by very different means, in Ref. \cite{Whalen17}. Inserting this result into Eqs.~(\ref{Lange0}), we finally arrive at the Langevin equation
\begin{equation}
\dot{z}(t)=L_{\text{S}}(z,z^{*})+f\left[\mathrm{e}^{\mathrm{i}\phi}z(t-T)-z(t)\right].\label{final_stoch_eq}
\end{equation}
Note that only when $\phi=0\,\,\mathrm{mod}\,\,2\pi$ do feedback and present-time oscillation interfere constructively, and we have positive, or in-phase, feedback. The opposite case, $\phi=\pi\,\,\mathrm{mod}\,\,2\pi$, is referred to as negative feedback.

\emph{Laser with delayed coherent feedback}.---To illustrate the reach of our description, we consider the application of delayed coherent feedback to a laser \cite{Lang80}. This is not only instructive from a pedagogical standpoint but also highly relevant in practice: the laser linewidth, which quantifies how close the emission is to monochromatic, can be dramatically narrowed via positive feedback, $\phi=0$, which we assume in what follows. In particular, it has been theoretically predicted \cite{Agrawal84} that, sufficiently above threshold, the narrowing scales as $(1+\kappa)^2$, where $\kappa\coloneqq{fT}$. This prediction has been confirmed by experiments achieving narrowing factors as large as $10^{6}$ \cite{Brunner17}, with a corresponding 3~Hz linewidth for a 194~THz laser. Our theory gives us access to parameter regimes not considered before. We find that an approximately polynomial narrowing, proportional to $(1+\kappa)^p$, holds for all laser parameters, but with a varying exponent $p$ that grows from $1$ well below the lasing threshold to $2$ well above it, and can exceed $2$ just above threshold---an effect that lies beyond the reach of linearized treatments.

We consider class-A lasers, in which the population and dipole lifetimes are much shorter than the cavity-photon lifetime, so that the material variables can be adiabatically eliminated. In the $P$ representation, the laser Liouvillian $\mathcal{L}_\text{S}$ translates into \cite{GardinerZollerBook,SupMat}
\begin{equation}
L_\text{S}(z,z^\ast)=\gamma\left(-1+\frac{C}{1+|z(t)|^2/n_0}\right)z(t)+g\sqrt{\gamma n_0}\eta(t),\label{laser_stoch_term}
\end{equation}
which involves just four real parameters: the cavity decay rate $\gamma$, the dimensionless pumping parameter $C$, the saturation photon number $n_0$, and the dimensionless quantum noise strength $g$. Here $\eta(t)$ is a complex Gaussian white noise with zero mean and $\overline{\eta(t)\eta^\ast(t')}=\delta(t-t')$ as the sole nonvanishing two-time correlator. Note that $\gamma$ and $n_0$ mathematically act as simple scale factors, so they need not be fixed in the simulations \cite{SupMat}.

Inserting (\ref{laser_stoch_term}) in (\ref{final_stoch_eq}) with $\phi=0$, we obtain the stochastic Langevin equation of the laser subject to positive delayed coherent feedback. Its deterministic part has two types of stable solutions: the trivial one, $z=0$, for $0<C<1$, and a family of nontrivial ones, $z=\sqrt{(C-1)n_0}\,\mathrm{e}^{\mathrm{i}\theta}$ with arbitrary $\theta$, for $C>1$. Hence, $C=1$ marks the laser threshold, above which classical emission sets in. Note that the laser model (\ref{laser_stoch_term})
assumes a photon number much smaller than its saturation value \cite{GardinerZollerBook}, i.e., $|z|^2\ll n_0$, which effectively limits the size of $C$ and $g$.

The laser emission is characterized by the power spectrum \cite{MandelWolfBook,Agrawal84}
\begin{equation}
S(\omega)\coloneqq\int_{-\infty}^{+\infty}\frac{{d}\tau}{2\pi}\mathrm{e}^{-\mathrm{i}\omega\tau}\lim_{t\rightarrow\infty}\overline{z(t)z^\ast(t+\tau)}.\label{power_spectrum}
\end{equation}
This power spectrum, written in terms of the intracavity field, is equal to that of the output field up to a multiplicative constant \cite{GardinerZollerBook,CNB-QOnotes}. In the absence of feedback and not too close to threshold, the two-time correlator is very well described by the exponential form $\lim_{t\rightarrow\infty}\overline{z(t)z^\ast(t+\tau)}=n_s \mathrm{e}^{-\frac12\Lambda|\tau|}$, with $n_s=\lim_{t\rightarrow\infty}\overline{|z(t)|^2}$ the steady-state mean photon number, leading to a Lorentzian spectrum of width $\Lambda$. Our simple Langevin equation (\ref{final_stoch_eq}) allows us to study the effect of feedback either numerically, by direct integration, or analytically, through suitable approximations. In general, feedback generates new resonances at approximately integer multiples of $2\pi/T$. We focus on the linewidth $\Lambda$ of the central resonance, at $\omega=0$. A linearized treatment of quantum fluctuations around the deterministic solutions, combined with the assumption $\Lambda T\ll1$, provides analytical insight \cite{SupMat}. Above threshold ($C>1$), we obtain $\Lambda=\Lambda_{\succ}/(1+\kappa)^2$, where $\Lambda_{\succ}=\frac12\gamma g^2/(C-1)$ is the linewidth in the absence of feedback, and we recall that $\kappa=fT$ is the product of the feedback strength and the delay. We thus recover the observed $(1+\kappa)^2$ narrowing, previously predicted through considerably more involved approaches \cite{Agrawal84,Brunner17}. Below threshold ($C<1$), we obtain $\Lambda=\Lambda_{\prec}/\sqrt{(1+\kappa)^2+\kappa\Lambda_{\prec}T/2}$, with $\Lambda_{\prec}=2\gamma(1-C)$. Consistency with $\Lambda T\ll1$ requires $(1+\kappa)^2\gg\kappa\Lambda_{\prec}T$, so that our linearized description yields an approximate $(1+\kappa)$ narrowing below threshold. In summary, the linearized theory predicts a crossover from a $(1+\kappa)$ to a $(1+\kappa)^2$ narrowing as the laser threshold is crossed.

\begin{figure}[t]
\includegraphics[width=1\columnwidth]{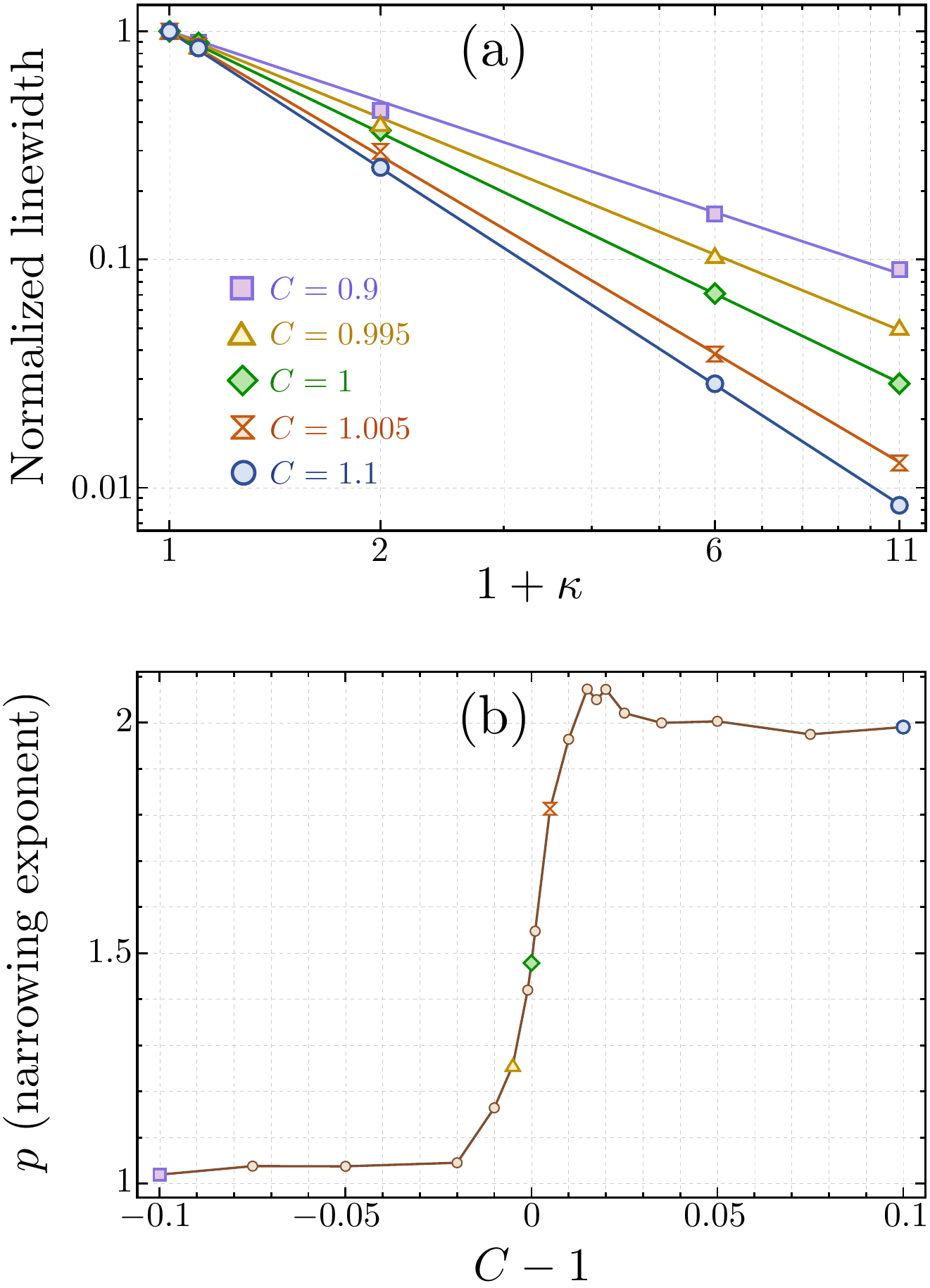}
\caption{Linewidth of a laser subject to delayed coherent feedback. (a) Laser linewidth, normalized to its feedback-free value, as a function of $1+\kappa=1+fT$ for different values of the pump parameter $C$, with the delay and noise strength fixed at $T=10\gamma^{-1}$ and $g=10^{-2}$, respectively. Symbols correspond to numerical simulations, and lines to fits of the power law $(1+\kappa)^{-p}$---note the log-log plot. (b) Narrowing exponent $p$ as a function of the distance to threshold $C-1$, for the same parameters as in (a). The exponent crosses over between the values $1$ (below threshold) and $2$ (above it) predicted by the linearized theory, and exhibits a maximum at $C-1\approx 0.015$--$0.02$, signaling enhanced narrowing just above threshold.}
\label{Fig_LaserResults}
\end{figure}

We confront these analytical predictions with numerical simulations of the full Langevin equation (\ref{final_stoch_eq}), for which efficient algorithms exist that allow us to explore the whole parameter space \cite{SupMat}. This again showcases the simplicity of our method compared with previous treatments \cite{Lang80,Agrawal84}. In Fig.~\ref{Fig_LaserResults}(a) we plot the laser linewidth as a function of $1+\kappa$ for different values of the pump $C\in[0.9,1.1]$, with the noise fixed to $g=10^{-2}$ and the delay to $T=10\gamma^{-1}$ (similar results are found for any other choice). For each $C$, we normalize the linewidth to its feedback-free ($f=0$) value and fit the data to a $(1+\kappa)^{-p}$ curve, a straight line of slope $-p$ in the log-log representation of the figure. In Fig.~\ref{Fig_LaserResults}(b) we plot the narrowing exponent $p$ as a function of $C-1$; it increases from $1$ to $2$ as the threshold is crossed. Interestingly, the increase is not perfectly monotonic: just above threshold, we find a region where $p>2$, i.e., where feedback narrows the linewidth more strongly than previously predicted. This is an effect that, to our knowledge, has not been reported before.

\emph{Conclusions.---}We have introduced a simple and general framework to describe open quantum systems coupled to non-Markovian environments, which contains the standard Markovian theory as a limiting case. Our approach uses normally ordered phase-space representations, such as the Glauber--Sudarshan $P$ representation or its positive variant due to Drummond and Gardiner, to map the system dynamics onto stochastic (Langevin) equations. The effect of the environment appears through two simple objects: (i) an additive stochastic driving force, which vanishes under the common assumption of an environment initially in the vacuum state, and (ii) a memory kernel that linearly couples the system to its own history. We have applied this framework to the paradigmatic problem of delayed coherent feedback, in which the system is driven by its own state at an earlier time. The simplicity of the method is evident in the resulting stochastic equations, which acquire precisely the terms one expects on classical grounds. As a practical demonstration, we have studied the linewidth of a laser subject to such feedback, reproducing experimental observations. Our analytical predictions, obtained within the standard linearized treatment, coincide with results previously derived from considerably more involved formalisms. Unlike previous approaches, ours further allows efficient numerical integration of the exact quantum dynamics with standard stochastic algorithms; in the laser case, these simulations reveal an unexpected linewidth narrowing just above threshold.

The simplicity and generality of our framework, together with the linear scaling of the number of stochastic equations with the number of bosonic modes, make it a powerful tool for the study of non-Markovian quantum dynamics from the few-body to the many-body regime. More importantly, by recasting the influence of a non-Markovian environment into a memory kernel acting on otherwise standard stochastic equations, our approach brings a broad class of open quantum systems within the scope of the well-established analytical and numerical machinery of stochastic calculus.

\begin{acknowledgments}
\textit{Acknowledgments}.---This work is part of project PID2023-153363NB-C22, funded by \seqsplit{MCIU/AEI/10.13039/501100011033}. C.N.-B. acknowledges sponsorship by the Generalitat Valenciana through CIDEGENT project \seqsplit{CIDEXG/2023/18}.
\end{acknowledgments}

\bibliography{NonMarkPandFeedbackLASER_Refs}

\onecolumngrid 

\newpage

\begin{center}
\Large\textbf{Supplemental material}
\end{center}

In this supplemental material we provide mathematical and technical details that have been omitted in the main text. In Section \ref{Sec:GlauberSudarshanP} we introduce the Glauber-Sudarshan $P$ representation and apply it to the system+environment model, obtaining the stochastic equations that we have manipulated in the main text. In Section \ref{Sec:Lambert} we provide physical arguments for the choice of memory kernel that we made for the delayed coherent feedback case. Finally, Section \ref{Sec:laser_solutions} provides a detailed description of the numerical approach that we used to solve the exact quantum dynamics of the laser subject to delayed coherent feedback, as well as the linearized approach capable of producing simple analytical expressions for its linewidth.

\section{Glauber-Sudarshan $P$ representation:\\stochastic equations for the system-environment model}\label{Sec:GlauberSudarshanP}

Here we introduce in detail the Glauber-Sudarshan representation of the system+environment problem presented in the first part of the main text, leading to the a set of stochastic equations representing the quantum dynamics. We start by defining the coherent states of the system and the environment in the usual way through $\hat{a}|\alpha\rangle=\alpha|\alpha\rangle$ and $\hat{b}_j|\beta_j\rangle=\beta_j|\beta_j\rangle$, with $\alpha\in\mathbb{C}$ and $\beta_j\in\mathbb{C}$, collecting all environmental amplitudes into the vector $\boldsymbol{\beta}=(...,\beta_{j-1},\beta_j,\beta_{j+1},...)$ for notational convenience. The combined state $\hat{\rho}$ of the system and the environment can be spanned in the operator-basis of coherent-state projectors $\hat{\Lambda}(\alpha,\boldsymbol{\beta})\coloneqq|\alpha\rangle\langle\alpha|\bigotimes_j|\beta_j\rangle\langle\beta_j|$ as \cite{Glauber,Sudarshan,CarmichaelBook} 
\begin{equation}
\hat{\rho}=\int d\mu(\alpha,\boldsymbol{\beta})P(\alpha,\boldsymbol{\beta})\hat{\Lambda}(\alpha,\boldsymbol{\beta}),\label{GS-PtoRho}
\end{equation}
where $P(\alpha,\boldsymbol{\beta})$ is the so-called Glauber-Sudarshan $P$ function and the integrals are defined over the whole complex plane of each variable, which defines the so-called phase space $\{\alpha,\boldsymbol{\beta}\}$, so that 
\begin{equation}
\int d\mu(\alpha,\boldsymbol{\beta})\coloneqq\int_\mathbb{C}d^2\alpha\prod_j\int_\mathbb{C}d^2\beta_j=\int_{\mathbb{R}^2}dXdY\prod_j\int_{\mathbb{R}^2}dX_jdY_j,
\end{equation}
where we have expressed the complex variables in terms of real and imaginary parts, $\alpha=X+\mathrm{i}Y$ and $\beta_j=X_j+\mathrm{i}Y_j$. As a concrete example, when the system and the environment are in the vacuum state $\hat{\rho}=|0\rangle\langle0|\bigotimes_j|0\rangle\langle0|$, the $P$ function takes the form $P(\alpha,\boldsymbol{\beta})=\delta^{(2)}(\alpha)\prod_j\delta^{(2)}(\beta_j)$ of a delta function centered at the origin of phase space. Hence, in such case, the phase space variables $\alpha=0=\beta_j$ are absent of fluctuations and bound to the phase space's origin. This fact has been used in the main text to set the environmental additive noise $F$ to zero.

Note that from (\ref{GS-PtoRho}) it is immediate to relate quantum expectation values in normal order with phase-space averages as
\begin{equation}
\langle\hat{a}^{\dagger m}\hat{a}^{n}\hat{b}_{j_1}^{\dagger m_1}\hat{b}_{j_1}^{n_1}...\hat{b}_{j_N}^{\dagger m_N}\hat{b}_{j_N}^{n_N}\rangle=\int d\mu(\alpha,\boldsymbol{\beta})P(\alpha,\boldsymbol{\beta})\alpha^{*m}\alpha^{n}\beta_{j_1}^{*m_1}\beta_{j_1}^{n_1}...\beta_{j_N}^{*m_N}\beta_{j_N}^{n_N}.
\end{equation}

Consider now the dynamical equation for the state $\hat{\rho}$:
\begin{equation}
\frac{d\hat{\rho}}{dt}=\Biggl[\frac{\hat{H}_\text{S}+\hat{H}_\text{E}+\hat{H}_\text{SE}}{\mathrm{i}\hbar},\hat{\rho}\Biggr]+\mathcal{L}_\text{S}[\hat{\rho}],\label{GenericMasterEq}
\end{equation}
where
\begin{subequations}
\begin{align}
\hat{H}_\text{S} & =\hbar\omega_\text{S}\hat{a}^\dagger\hat{a},
\\
\hat{H}_\text{E} & =\sum_j\hbar\omega_j\hat{b}_j^\dagger\hat{b}_j,
\\
\hat{H}_\text{SE} & =\sum_j\hbar(g_j^\ast\hat{b}_j^\dagger\hat{a}+g_j\hat{b}_j\hat{a}^\dagger),
\end{align}
\end{subequations}
and $\mathcal{L}_\text{S}[\hat{\rho}]$ is a generic Liouvillian that can account for other processes contributing to the dynamics of the system, as mentioned in the main text. Let
us drop, momentarily, this generic Liouvillian and focus on the part accounting for the S-E interaction. Using the identities
\begin{subequations}\label{Pidentities}
\begin{align}
\hat{a}\hat{\Lambda} & =\alpha\hat{\Lambda},\;\;\;\hat{\Lambda}\hat{a}=(\alpha+\partial_{\alpha^\ast})\hat{\Lambda},
\\
\hat{\Lambda}\hat{a}^\dagger & =\alpha^\ast\hat{\Lambda},\;\;\;\hat{a}^\dagger\hat{\Lambda}=(\alpha^\ast+\partial_\alpha)\hat{\Lambda},
\\
\hat{b}_j\hat{\Lambda} & =\beta_j\hat{\Lambda},\;\;\;\hat{\Lambda}\hat{b}_j=(\beta_j+\partial_{\beta_j^\ast})\hat{\Lambda},
\\
\hat{\Lambda}\hat{b}_j^\dagger & =\beta_j^\ast\hat{\Lambda},\;\;\;\hat{b}_j^\dagger\hat{\Lambda}=(\beta_j^\ast+\partial_{\beta_j})\hat{\Lambda},
\end{align}
\end{subequations}
and partial integration to bring the derivatives from $\hat{\Lambda}(\alpha,\boldsymbol{\beta})$ to $P(\alpha,\boldsymbol{\beta})$ under the usual physical assumption that $P(\alpha,\boldsymbol{\beta})$ falls to zero fast enough at the boundaries of phase space, (\ref{GenericMasterEq}) can be turned into a partial differential equation for the $P$ function
that reads 
\begin{equation}
\partial_{t}P(\alpha,\boldsymbol{\beta};t)=-\left[\partial_\alpha A_\alpha+\partial_{\alpha^\ast}A_{\alpha^\ast}+\sum_j\left(\partial_{\beta_j}A_{\beta_j}+\partial_{\beta_j^\ast}A_{\beta_j^\ast}\right)\right]P(\alpha,\boldsymbol{\beta};t),\label{P-FP}
\end{equation}
with so-called drift vector components\begin{subequations}
\begin{align}
A_\alpha & =-\mathrm{i}\omega_\text{S}\alpha-\mathrm{i}\sum_j g_j\beta_j,\qquad A_{\alpha^\ast}=A_\alpha^\ast,
\\
A_{\beta_j} & =-\mathrm{i}\omega_j\beta_j-\mathrm{i}g_j^\ast\alpha,\qquad A_{\beta_j^\ast}=A_{\beta_j}^\ast.
\end{align}
\end{subequations}
Eq. (\ref{P-FP}) is a Fokker-Planck equation (without diffusion, so-called Liouville equation), which according to Ito rules \cite{CarmichaelBook,GardinerBook} is equivalent to
the following set of stochastic Langevin equations
\begin{subequations}\label{LangeP_SI}
\begin{align}
\dot{\alpha} & =A_\alpha=-\mathrm{i}\omega_\text{S}\alpha-\mathrm{i}\sum_j{g}_j\beta_j,\label{dalpha0app}
\\
\dot{\beta}_j & =A_{\beta_j}=-\mathrm{i}\omega_j\beta_j-\mathrm{i}{g}_j^\ast\alpha,\label{dbetaApp}
\end{align}
\end{subequations}
where the equivalence must be understood in the sense that the statistics derived from the phase-space quasi-distribution $P(\alpha,\boldsymbol{\beta};t)$ equal stochastic averages, that
is, for any phase space function $G(\alpha,\boldsymbol{\beta})$ we have
\begin{equation}
\int d\mu(\alpha,\boldsymbol{\beta})P(\alpha,\boldsymbol{\beta};t)G(\alpha,\boldsymbol{\beta})=\overline{G(\alpha(t),\boldsymbol{\beta}(t))},
\end{equation}
where the bar denotes stochastic average using the solution of the stochastic equations (\ref{LangeP_SI}). Note that due to the simple form of the Hamiltonian, these stochastic equations have no noise other than in the initial conditions of the stochastic variables $\alpha$ and $\beta_j$. When considering the rest of processes acting on the system and included in $\mathcal{L}_\text{S}[\hat{\rho}]$, new terms might appear on the Eq. (\ref{dalpha0app}) that might include noise as well. However, remarkably, these terms which depend solely on system operators do not change the Eq. (\ref{dbetaApp}) for the environmental amplitudes $\beta_j$.

\section{Physical arguments for the choice of $K(\tau)$ \\for the delayed coherent feedback problem}\label{Sec:Lambert}

As discussed in the main text, for a delay $T$, the kernel $K(\tau)$ must be peaked both at $\tau=T$ and at $\tau=-T$, owing to its Hermiticity. Approximating this behavior by Dirac deltas, we initially write $K(\tau)=-f[e^{\mathrm{i}\phi}\delta(\tau-T)+e^{-\mathrm{i}\phi}\delta(\tau+T)]$, with $f$ a real and positive constant with dimensions of frequency and $\phi$ an arbitrary phase; the minus sign is written for later convenience. The kernel also needs to satisfy the condition $K(0)\ge|K(\tau)|$ $\forall\tau$, which requires adding another peak at $\tau=0$ with a weight larger than or equal to $f$. Let us then write
\begin{equation}
K(\tau)=f\left[2\varepsilon\delta(\tau)-e^{\mathrm{i}\phi}\delta(\tau-T)-e^{-\mathrm{i}\phi}\delta(\tau+T)\right].
\end{equation}
with $\varepsilon\geq1/2$. Next, we show that $\varepsilon=1$ must be chosen for physical consistency.

Assuming an environment that starts in the vacuum state, the Langevin equation for the slowly-varying interaction-picture amplitude takes the form {[}see Eq. (3) in the main text{]}: 
\begin{equation}
\dot{z}(t)=-\int_{0}^{t}d\tau K(\tau)z(t-\tau)=-\varepsilon fz(t)+f\mathrm{e}^{\mathrm{i}\phi}z(t-T).\label{Langevin_no}
\end{equation}
The rationale behind the need for $\varepsilon=1$ follows from understanding that, in the positive feedback case ($\phi=0$), the field lost by S toward E is returned intact (with the same phase) by E at a later time. So, if S is initially in an arbitrary coherent state, that state should remain forever. This situation corresponds to an arbitrary constant solution to Eq. (\ref{Langevin_no}). We prove next that this is only possible if $\varepsilon=1$.

Eq. (\ref{Langevin_no}) is a homogeneous linear differential equation with delay. In the absence of delay ($T=0$) the (homogeneous) solution of Eq. (\ref{Langevin_no}) is proportional to the exponential $z(t)\sim\mathrm{e}^{\lambda t}$, with $\lambda=f(\mathrm{e}^{\mathrm{i}\phi}-\varepsilon)$. Already here, without delay, we see that if $\varepsilon>1$, $\text{Re}\{\lambda\}<0$ and the only long time term solution is the trivial one, $\lim_{t\rightarrow\infty}z(t)=0$; in contrast, for $\varepsilon=1$, $\text{Re}\{\lambda\}=0$, allowing for nontrivial solutions, which are furthermore constant for $\phi=0$ ($\lambda=0$). The effect of the delay, even if small, is qualitatively huge. As we shall see, for $T>0$ there are an infinite number of exponentials that solve the homogeneous problem. Plugging now $z(t)\sim\mathrm{e}^{\lambda t}$ into Eq. (\ref{Langevin_no}) yields a characteristic equation that can be written as, 
\begin{equation}
\left(\lambda+\varepsilon f\right)=f\mathrm{e}^{\mathrm{i}\phi}\mathrm{e}^{-\lambda T}.\label{char}
\end{equation}
Multiplying both sides by $T\mathrm{e}^{(\lambda+\varepsilon)T}$ and defining $w\coloneqq\left(\lambda+\varepsilon f\right)T$ and $x\coloneqq fT\mathrm{e}^{\varepsilon fT}\mathrm{e}^{\mathrm{i}\phi}$, Eq. (\ref{char}) can be written as $w\mathrm{e}^{w}=x$, which is the equation satisfied by the Lambert W function \cite{Corless96}:
$w=W(x)$. The Lambert W function has an infinite number of solutions or branches $\{W_k(x)\}_{k\in\mathbb{Z}}$. $W_0$ is known as the principal branch and verifies $W_0(x\mathrm{e}^{x})=x$. For real $x>-1/e$, $W_0$ is the only branch that evaluates to a real number, while the rest of branches verify $W_{-k}(-x)=W_k^\ast(x)$.
Therefore, the solutions to (\ref{char}) are given by 
\begin{equation}
\lambda_kT=-\varepsilon{g}+W_k({g}\mathrm{e}^{\varepsilon g}\mathrm{e}^{\mathrm{i}\phi}),\qquad{g}\coloneqq fT\ge0.\label{lam_gen}
\end{equation}

In the special case $\phi=0$ (positive feedback) and $\varepsilon=1$, we have $W_0(g\mathrm{e}^g)=g$, so $\lambda_0=0$. The analysis of the properties of the Lambert W function is not trivial \cite{Corless96}. However it is easy to convince oneself, e.g. by a numerical study of Eq. (\ref{lam_gen}) that $\text{Re}\{\lambda_0\}>\text{Re}\{\lambda_k\}$ $\forall k\ne0$, and $\text{Re}\{\lambda_0\}<0$ for $\phi\ne0$ or $\varepsilon>1$. From here we conclude two main results: (i) the stability of Eq. (\ref{Langevin_no}) is guaranteed as $\text{Re}\{\lambda_k\}\le0$ $\forall k$, and (ii) only for $\varepsilon=1$ the positive feedback case admits an arbitrary constant solution, representing an arbitrary coherent state. In any other case the asymptotic solution in the absence of noise decays to zero.

\section{the laser: numerical and linearized analysis}\label{Sec:laser_solutions}

In the main text we studied class-A lasers subject to positive delayed coherent feedback. Using the stochastic Langevin equations derived for these lasers in Eq. (9.3.78) of \cite{GardinerZollerBook} for the feedback-free case, and including feedback within our formalism, we obtain 
\begin{equation}
\dot{z}(t)=A(z)+fz(t-T)+g\sqrt{\gamma n_0}\eta(t),\label{full_stoch_eq}
\end{equation}
with
\begin{align}
A(z) & =\left(-\gamma-f+\frac{\gamma C}{1+|z(t)|^2/n_0}\right)z(t),
\end{align}
from which we ought to analyze the laser's linewidth, defined in the literature \cite{MandelWolfBook,Agrawal84} as the width of the central peak of the power spectrum
\begin{equation}
S(\omega)=\int_{-\infty}^{+\infty}\frac{d\tau}{2\pi}\mathrm{e}^{-\mathrm{i}\omega\tau}\lim_{t\rightarrow\infty}\overline{z(t)z^\ast(t+\tau)}.\label{PowerSpectrum_Appendix}
\end{equation}

It is worth mentioning that in Eq. (9.3.78) of \cite{GardinerZollerBook} the constant multiplying the noise is written differently. Here we have chosen the form $g\sqrt{\gamma n_0}$ for that constant because then both $\gamma$ and $n_0$ act as scaling factors only, so effectively $C$ and $g$ are the only free parameters of the laser, with the latter characterizing the size of quantum noise. In order to see this explicitly, note that in terms of the dimensionless time $\tilde{t}=\gamma t$ and the normalized variable $\tilde{z}(\tilde{t})=z(t)/\sqrt{n_0}$, Eq. (\ref{full_stoch_eq}) takes the form
\begin{equation}
\frac{d\tilde{z}}{d\tilde{t}}=\left(-1-\tilde{f}+\frac{C}{1+|\tilde{z}(\tilde{t})|^2}\right)\tilde{z}(\tilde{t})+\tilde{f}\tilde{z}(\tilde{t}-\tilde{T})+g\tilde{\eta}(\tilde{t}),\label{full_stoch_eq_normalized}
\end{equation}
where $\tilde{f}=f/\gamma$, $\tilde{T}=\gamma T$, and $\tilde{\eta}(\tilde{t})=\eta(t)/\sqrt{\gamma}$ is a complex Gaussian noise with zero mean and $\overline{\tilde{\eta}(\tilde{t})\tilde{\eta}^*(\tilde{t}')}=\delta(\tilde{t}-\tilde{t}')$ as its only non-vanishing two-time correlator. These normalizations clearly show that $g$ effectively controls the size of the noise, $\gamma$ and $\gamma^{-1}$ act as simple normalizing scales for rates and times, respectively, and $n_0$ is just a scaling factor for the variable $z$.

\subsection{Numerical simulations}

In order to solve numerically Eq. (\ref{full_stoch_eq}) we use the semi--implicit algorithm developed in \cite{Drummond91} (see also \cite{Drummond24}). This algorithm is a finite-differences based method in which the total integration time $t_\text{end}$ is divided in $N$ segments, creating hence a time grid $\{t_n=n\Delta t\}_{n=0,1,...,N}$, with time step $\Delta t=t_\text{end}/N$ and starting at $t=0$. Special care must be put in this problem with feedback to choose the time step $\Delta t$ and the delay $T$ such that they have an integer relation, say $T=N_T\Delta t$ with $N_T$ integer. We assume that the feedback vanishes before $t=0$, that is, $z(t)=0$ for $t<0$. Then, a recursive algorithm starts in which the amplitudes at time
$t_n$, $z_n\coloneqq z(t_n)$, are found from the amplitudes $z_{n-1}$ at an earlier time $t_{n-1}$ from the rule 
\begin{equation}\label{stochastic_finite_differences}
z_n=z_{n-1}+\Delta tf\tilde{z}_{n-N_T}+\Delta t A(\tilde{z}_n)+g\sqrt{\gamma n_0}\frac{W_{n,\text{R}}+\mathrm{i}W_{n,\text{I}}}{\sqrt{2}},\;\;\;n=1,2,...,N.
\end{equation}
$\{W_{n,\sigma}\}_{n=1,2,...,N}^{\sigma=\text{R},\text{I}}$ are independent discrete noises (Wiener increments \cite{GardinerBook,CarmichaelBook}) drawn from normal distributions with zero mean and variance $\Delta t$. $\tilde{z}_n$ is an approximation to the variables at the middle of the $\{t_{n-1},t_n\}$ interval (hence the name ``semi-implicit'' for the algorithm), found from the following iterative algorithm 
\begin{equation}
\tilde{z}_n^{(p)}=z_{n-1}+\frac{1}{2}\left[\Delta tf\tilde{z}_{n-N_T}+\Delta tA(\tilde{z}_n^{(p-1)})+g\sqrt{\gamma n_0}\frac{W_{n,\text{R}}^{(p)}+\mathrm{i}W_{n,\text{I}}^{(p)}}{\sqrt{2}}\right],
\end{equation}
where the iterative index starts at $p=1$ with $\tilde{z}_n^{(0)}=z_{n-1}$, and two iterations are typically carried in all our simulations, that is, $\tilde{z}_n=\tilde{z}_n^{(2)}$. Note that the Wiener increments $W_{n,\sigma}^{(p)}$ are independent from each other and from the ones in Eq. (\ref{stochastic_finite_differences}), that is, they are each drawn from a normal distribution with zero mean and variance $\Delta t$.

The sequence $\{z_n\}_{n=0,1,...,N}$ is obtained for many stochastic realizations (typically between $10^5$ and $10^6$), so-called \emph{stochastic trajectories}. We obtain an approximation to the correlation function $\lim_{t\rightarrow\infty}\overline{z(t)z^\ast(t+\tau)}$ required for the power spectrum (\ref{PowerSpectrum_Appendix}) as follows. First, we choose a sufficiently large time $t_\infty=n_\infty\Delta t$ that will play the role of $t\rightarrow\infty$. For each trajectory, we then compute the sequence $\{z_{n_\infty}z^\ast_{n_\infty+m}\}_{m=-m_\text{max},-m_\text{max}+1,...,m_\text{max}}$, where $m_\text{max}=N-n_\infty$ and average them. The result provides a discrete approximation to the correlation function, that is, $\big\{\lim_{t\rightarrow\infty}\overline{z(t)z^\ast(t+\tau_m)}\big\}_{m=-m_\text{max},-m_\text{max}+1,...,m_\text{max}}$, with $\tau_m=m\Delta t$. Using then a standard FFT algorithm we evaluate the desired power spectrum (\ref{PowerSpectrum_Appendix}), whose convergence as a function of $n_\infty$, $m_\text{max}$, $\Delta t$ and the number of stochastic trajectories we check.

\subsection{Linearized analysis below threshold}

Below threshold, where the deterministic solution is 0, we obtain a linearized description with respect to quantum fluctuations by assuming that $z(t)$ is small. Keeping only linear terms in Eq. (\ref{full_stoch_eq}), we obtain the linear stochastic equation
\begin{equation}
z(t)=-\gamma(1-C)z(t)+f[z(t-T)-z(t)]+g\sqrt{\gamma n_0}\eta(t).\label{lin_below_th}
\end{equation}
This equation can be solved employing common Fourier transform techniques. In particular, defining the Fourier transform of any function $F(t)$ as $\tilde{F}(\omega)=\int_{-\infty}^{+\infty}\frac{dt}{2\pi}e^{-\mathrm{i}\omega t}F(t)$, with inverse relation $F(t)=\int_{-\infty}^{+\infty}dte^{\mathrm{i}\omega t}\tilde{F}(\omega)$, we find
\begin{equation}
\tilde{b}(\omega)=\mathcal{A}(\omega)\tilde{\eta}(\omega),\hspace{3mm}\text{with }\mathcal{A}(\omega)=\frac{g\sqrt{\gamma n_0}/2\pi}{\mathrm{i}\omega+\gamma(1-C)+f\left(1-e^{-\mathrm{i}\omega T}\right)},\label{b_below_th}
\end{equation}
and where the transformed noise $\tilde{\eta}(\omega)$ has zero mean and spectral correlator $\overline{\tilde{\eta}(\omega)\tilde{\eta}^\ast(\omega')}=\delta(\omega-\omega')/2\pi$, with any other second-order correlator vanishing. Inserting this solution into (\ref{PowerSpectrum_Appendix}) we find
\begin{equation}
S(\omega)=\frac{|\mathcal{A}(\omega)|^2}{2\pi}=\frac{g^2\gamma n_0/2\pi}{\omega^2+f^2+[f+\gamma(1-C)]^2-2f[f+\gamma(1-C)]\cos[\omega T]-2f\omega\sin[\omega T]}.
\end{equation}
Since we are interested in the central resonance, we next assume $\omega T\ll1$ and expand the denominator to second order on $\omega$, obtaining
\begin{equation}
S(\omega)\approx\frac{g^2\gamma n_0/2\pi}{\omega^2[(1+\kappa)^2+\kappa T\gamma(1-C)]+\gamma^2(1-C)^2},
\end{equation}
which has a Lorentzian shape $S_0/\left(\omega^2+\Lambda^2/4\right)$ with full width at half maximum $\Lambda=2\gamma(1-C)/\sqrt{(1+\kappa)^2+\kappa T\gamma(1-C)}$. This is the analytic prediction for the linewidth below threshold that we provided in the main text.

\subsection{Linearized analysis above threshold}

Linearization is way more subtle above threshold \cite{CNB08,CNB10,CNB17}. As the deterministic terms do not fix the phase of $z(t)$ in the steady state, quantum noise will be able to act unboundedly on it, so fluctuations around a given deterministic solution with given phase cannot be considered small. In principle, in such cases we need to use a generalized linearized theory suitable for systems with spontaneous symmetry breaking of a continuous symmetry \cite{CNB08,CNB10,CNB17}. However, in the case of the laser it is easy to see that this generalized approach is equivalent to an expansion $z(t)=\Big[\sqrt{n_0(C-1)}+b(t)\Big]\mathrm{e}^{\mathrm{i}\theta(t)}$, where the phase $\theta(t)$ is taken as a fully dynamical stochastic variable that need not be small, while $b(t)$ is real and provides the amplitude fluctuations, which can be assumed small. We can then linearize Eq. (\ref{full_stoch_eq}) with respect to $b(t)$ and $\dot{\theta}(t)$,
obtaining two uncoupled linear equations
\begin{subequations}
\begin{align}
\dot{\theta} & =f[\theta(t-T)-\theta(t)]+g\sqrt{\frac{\gamma/2}{C-1}}\eta_{\theta}(t),\label{dtheta}
\\
\dot{b}(t) & =-2\gamma\left(1-C^{-1}\right)b(t)+f[b(t-T)-b(t)]+g\sqrt{\frac{\gamma n_0}{2}}\eta_{b}(t),\label{db}
\end{align}
\end{subequations}
where additionally we have assumed a small enough phase diffusion over a delay time $T$, such that we can also linearize with respect to $[\theta(t-T)-\theta(t)]$. The noises $\eta_{b}(t)=\sqrt{2}\text{Re}\{\eta(t)\}$ and $\eta_{\theta}(t)=\sqrt{2}\text{Im}\{\eta(t)\}$, are real and independent, that is, $\overline{\eta_j(t)\eta_l(t')}=\delta_{jl}\delta(t-t')$.

These equations can be used to evaluate the two-time correlator in the power spectrum (\ref{PowerSpectrum_Appendix}). Specifically, we get
\begin{equation}
\lim_{t\to\infty}\overline{z(t)z^\ast(t+\tau)}=\lim_{t\to\infty}\left[n_0(C-1)+\overline{b(t)b^\ast(t+\tau)}\right]\;\overline{e^{\mathrm{i}[\theta(t)-\theta(t+\tau)]}},\label{TwoTimeCorr_Appendix}
\end{equation}
where we have used the fact that $b$ and $\theta$ are uncorrelated, and that the amplitude fluctuations average to zero in the long term, $\lim_{t\to\infty}\overline{b(t)}=0$. This expression can be simplified further by noting two things. First, consistency with the linearization assumptions requires amplitude fluctuations to be much smaller than the deterministic amplitude, that is, $\lim_{t\to\infty}\overline{b(t)b^\ast(t+\tau)}\ll n_0(C-1)$. Sufficiently above threshold this is ensured by the fact that Eq. (\ref{db}) has damping at rate $2\gamma\left(1-C^{-1}\right)$, which, in combination with a small $g$, bounds the amplitude fluctuations to moderate values. The second thing we note is that $\theta(t)$ is a zero-mean
Gaussian stochastic variable (within this linearized theory), and therefore \cite{MandelWolfBook} $\overline{e^{\mathrm{i}[\theta(t)-\theta(t')]}}=e^{-\overline{[\theta(t)-\theta(t')]^2}/2}$.
With these considerations, (\ref{TwoTimeCorr_Appendix}) is simplified to
\begin{equation}
\lim_{t\to\infty}\overline{z(t)z^\ast(t+\tau)}=n_0(C-1)e^{-V_{\theta}(\tau)/2},\label{TwoTimeCorrSimplified}
\end{equation}
where we have defined the phase variance
\begin{equation}
V_{\theta}(\tau)=\lim_{t\to\infty}\overline{[\theta(t)-\theta(t+\tau)]^2}.\label{Vtheta}
\end{equation}
It is not difficult to evaluate this variance from (\ref{dtheta}). Employing common Fourier techniques, one obtains
\begin{equation}
\theta(t)=g\sqrt{\frac{\gamma}{2(C-1)}}\int_{-\infty}^{+\infty}d\omega\underbrace{\frac{1}{\mathrm{i}\omega+f\left(1-e^{-\mathrm{i}\omega T}\right)}}_{H(\omega)}e^{\mathrm{i}\omega t}\tilde{\eta}_{\theta}(\omega),
\end{equation}
where the transformed noise $\tilde{\eta}_{\theta}(\omega)$ has zero mean and spectral correlation $\overline{\tilde{\eta}_{\theta}(\omega)\tilde{\eta}_{\theta}(\omega')}=\delta(\omega+\omega')/2\pi$. We thus obtain
\begin{equation}
\theta(t)-\theta(t+\tau)=g\sqrt{\frac{\gamma}{2(C-1)}}\int_{-\infty}^{+\infty}d\omega H(\omega)\left(1-e^{\mathrm{i}\omega\tau}\right)e^{\mathrm{i}\omega t}\tilde{\eta}_{\theta}(\omega),
\end{equation}
which plugged into (\ref{Vtheta}) leads to
\begin{align}
V_{\theta}(\tau) & =\frac{g^2\gamma}{2(C-1)}\int_{-\infty}^{+\infty}d\omega\int_{-\infty}^{+\infty}d\omega'H(\omega)H(\omega')\left(1-e^{\mathrm{i}\omega\tau}\right)\left(1-e^{\mathrm{i}\omega'\tau}\right)e^{\mathrm{i}(\omega+\omega')t}\overline{\tilde{\eta}_{\theta}(\omega)\tilde{\eta}_{\theta}(\omega')}
\\
 & =\frac{g^2\gamma}{4\pi(C-1)}\int_{-\infty}^{+\infty}d\omega\int_{-\infty}^{+\infty}d\omega'\left|H(\omega)\right|^2\left|1-e^{\mathrm{i}\omega\tau}\right|^2\nonumber 
 \\
 & =\frac{g^2\gamma/\pi}{C-1}\int_{-\infty}^{+\infty}d\omega\frac{\sin^2(\omega\tau/2)}{\omega^2+2f\omega\sin(\omega T)+2f^2[1-\cos(\omega T)]},\nonumber 
\end{align}
where we have used $H^\ast(\omega)=H(-\omega)$. If we focus on the central resonance, as we did below threshold, then the denominator can be expanded up to second order in $\omega T$, leading to
\begin{equation}
V_{\theta}(\tau)\approx\frac{g^2\gamma/\pi}{(C-1)(1+2fT+f^2T^2)}\underbrace{\int_{-\infty}^{+\infty}d\omega\frac{\sin^2(\omega\tau/2)}{\omega^2}}_{\pi|\tau|/2}=\underbrace{\frac{g^2\gamma}{2(C-1)(1+\kappa)^2}}_{\coloneqq\Lambda}|\tau|.
\end{equation}
Hence, the two time correlator (\ref{TwoTimeCorrSimplified}) gets an exponential form 
\begin{equation}
\lim_{t\to\infty}\overline{z(t)z^\ast(t+\tau)}=n_0(C-1)e^{-\Lambda|\tau|/2},
\end{equation}
leading to a Lorentzian power spectrum (\ref{PowerSpectrum_Appendix}) of width $\Lambda$, which is the analytic prediction that we put forward in the main text for the linewidth above-threshold.
\end{document}